\documentclass[prb,twocolumn,amsmath,amssymb,showpacs,superscriptaddress]{revtex4}
\usepackage{graphicx}% Include figure files
\usepackage{dcolumn}% Align table columns on decimal point
\usepackage{bm}% bold math

\newcommand{\ö}{$\ddot{\mathrm{o}}$}

\begin{document}

\title{Two Component Heat Diffusion Observed in CMR Manganites}% by Femto Second Spectroscopy}%

\author{J. Bielecki}%
\email{johan.bielecki@chalmers.se}
\affiliation{%
Department of Applied Physics, Chalmers University of Technology, SE-41296 G$\ddot{o}$teborg Sweden }
\author{R. Rauer}
\affiliation{%
Department of Applied Physics, Chalmers University of Technology, SE-41296 G$\ddot{o}$teborg Sweden }
\author{E. Zanghellini}
\affiliation{%
Department of Applied Physics, Chalmers University of Technology, SE-41296 G$\ddot{o}$teborg Sweden }
\author{R. Gunnarsson}
\affiliation{%
Department of Microtechnology and Nanoscience (MC2), Chalmers University of Technology, SE-412 96 G$\ddot{o}$teborg, Sweden}
\author{K. D{\ö}rr}
\affiliation{%
Institut F$\ddot{u}$r Festk$\ddot{o}$rper- und Werkstofforschnung, D-01069 Dresden Germany}
\author{L. B{\ö}rjesson}
\affiliation{%
Department of Applied Physics, Chalmers University of Technology, SE-41296 G$\ddot{o}$teborg Sweden }

\date{\today}

\begin{abstract}
We investigate the low-temperature electron, lattice, and spin dynamics of $\rm{LaMnO}_3$ (LMO) and $\rm{La}_{0.7}\rm{Ca}_{0.3}\rm{MnO}_3$ (LCMO) by resonant pump-probe reflectance spectroscopy. Probing the high-spin d-d transition as a function of time delay and probe energy, we compare the responses of the Mott insulator and the double-exchange metal to the photoexcitation. Attempts have previously been made to describe the sub-picosecond dynamics of CMR manganites in terms of a phenomenological three temperature model describing the energy transfer between the electron, lattice and spin subsystems followed by a comparatively slow exponential decay back to the ground state. However, conflicting results have been reported. Here we first show clear evidence of an additional component in the long term relaxation due to film-to-substrate heat diffusion and then develop a modified three temperature model that gives a consistent account for this feature.   We confirm our interpretation by using it to deduce the bandgap in LMO. In addition we also model the non-thermal sub-picosecond dynamics, giving a full account of all observed transient features both in the insulating LMO and the metallic LCMO. %We conclude with an example of how our new model can be used to measure the bandgap in the semiconducting LMO.
\end{abstract}


\maketitle
\section{Introduction}
Transition-metal oxides with partially filled d-electron bands exhibit a wide range of fascinating phenomena like dynamic lattice distortions, magnetic, charge, and orbital ordering. Particularly the perovskite manganese oxides $\rm{La}_{(1-x)}\rm{A}_x\rm{MnO}_3$, where A is a divalent alkali metal, have attracted scientific attention. While the undoped $\rm{LaMnO}_3$ (LMO) is an A-type antiferromagnetic Mott insulator \cite{Wollan,Kovaleva} of orthorhombic symmetry, $\rm{La}_{0.7}\rm{Ca}_{0.3}\rm{MnO}_3$ (LCMO) exhibits the colossal magnetoresistance (CMR) effect along with the transition from a paramagnetic insulating to a ferromagnetic metallic phase \cite{Schiffer}. It is accompanied by a structural transition from orthorhombic to rhombohedral symmetry. With respect to technical applications of the CMR manganites, for instance in magneto-optical switches, the dynamics of their strongly correlated electron, lattice, and spin subsystems are of particular interest \cite{Millis}.

Pump-probe spectroscopy provides a powerful technique to investigate these dynamics: A short fs pump-laser pulse yields instantaneous photoexcitation of the electron system via electric-dipole allowed transitions. It has been established that, after the restoration of a Fermi distribution by electron-electron interactions ($< 1$ ps), excess heat is transferred from the electron system to the lattice on a ps time scale via phonon emission \cite{OgasawaraGeneral}. Successively, heat is transferred from the lattice to the spin system on a ps to ns time scale mediated by the spin-orbit interaction \cite{OgasawaraKerr, OgasawaraGeneral, Beaurepaire}. The dynamics of the electron, phonon, and spin systems are probed by a second laser pulse delayed with respect to the pump pulse.

Previous pump-probe studies of the CMR manganites have focused on the spin-lattice (l-s) relaxation while the dynamics of the electron-electron (e-e) and electron-lattice (e-l) interactions have been largely neglected so far. For instance, the temperature dependence of the transient transmission of thin LSMO and LCMO films was analyzed in terms of two components with time scales $<1$ ps and 20--200 ps. While the sub-ps transient is naturally associated with e-e processes, the slower component was attributed to the l-s relaxation \cite{Lobad}. %The spin-lattice relaxation time ($\tau_{ls}$) was shown to follow the magnetic specific heat implying a constant spin-lattice coupling strength with temperature.

Recently however, differential transmission spectroscopy studies on the antiferromagnetic parent compound LMO explained the neglected intermediate transients in terms of the relaxation of the collective Jahn-Teller distortion and energy transfer between the electron and phonon subsystems. Curiously though, the l-s component seemed absent \cite{Tamaru}. Similarly, the corresponding relaxation in LCMO was attributed to Jahn-Teller polarons \cite{Wu}. There are, however, features in the picosecond dynamics in CMR manganites not covered in the published literature. One such feature is a sign change after several hundreds of picoseconds \cite{Hirobe}. With the spin-lattice relaxation time in the range 10-100 ps, such a zero crossing is not possible according to the models previously used. Also, the observed dependence of the heat diffusion rate on the probe energy\cite{OgasawaraKerr} is in contrast to the models used.

In this paper, we study the free carrier, electron, lattice, and spin dynamics of LMO and LCMO by resonant pump-probe reflectance spectroscopy. A modification of the widely used phenomenological three temperature model first described by Beaurepaire \cite{Beaurepaire} is developed in order to explain all the observed features, including the aforementioned features previously unaccounted for, in both LMO and LCMO. For delays larger than 1 ps we observe three clearly distinguishable components in the low-temperature transients. First we see a fast component reflecting the transient electron population after photoexcitation ($\sim 5$ ps) followed by an intermediate component reflecting the lattice temperature dynamics, exponentially rising with the e-l relaxation time, exponentially decaying with the l-s relaxation time. Subsequently we have a slow component (exponential rise time $\sim 80$ ps) in agreement with previous reports of conventional l-s heating by spin-orbit interaction. We conclude that the intermediate component relates to the presence of exchange mechanisms in the ground state. In CMR compounds, lattice heating induces a relaxation of the Jahn-Teller distortion which causes a change in the transfer matrix of the exchange mechanisms. Following these relaxation components we observe finally a decay on the ns scale consisting of two components. This bi-exponential decay back to the ground state has been mentioned before\cite{OgasawaraKerr} but without much discussion. Here we show a clear separation of the two processes where the faster of these is interpreted as the film-to-substrate heat diffusion ($\sim 500$ps) while the slower ($\sim 10$ ns) is interpreted as the in-film heat diffusion.

  The sub-picosecond response due to changes in the density of states immediately following the photoexcitation is seen as a sharp peak centered around zero delay, and the rise and decay of this peak is modeled by a phenomenological differential equation that almost exactly reproduces the observed response. Solving the differential equation, the response is basically a convolution between the pump pulse and an exponential decay ($\sim 70$fs) representing the thermalization of the excited electrons back to a well defined fermi level. There is also signs of a photoinduced phase transition in LMO where the $\sim 70$fs component of metallic origin is accompanied by a slower ($\sim 850$ fs) component indicating the reformation of the band gap. Using the three temperature model modified with two component heat diffusion together with the modeled non-thermal response, all observed features in the transients of CMR manganites can be understood.

\section{Experimental}
The transient reflectivity measurement was performed at near normal incidence with a Ti:sapphire mode-locked amplifier with a 1kHz repetition rate and nominal $170$ fs pulse length. The pump beam uses the fundamental at 775 nm while the probe energy $E_{pr}$ is tuned over the range 0.59--2.25 eV using an optical parametric amplifier (TOPAS). The beam diameters at the sample surface was adjusted to $135$ and $160$ $\mu \rm{m}$ for the probe and pump respectively, giving a constant pump fluence $< 1 \rm{mJ}/\rm{cm}^2$ and a probe/pump fluence ratio of less than $1/30$. Using the cross-correlation between pump and probe, the effective time resolution is estimated to $\sim 190$ fs, while autocorrelation gives an individual pulse length close to the nominal 170 fs.

The LMO and LCMO films of respective thicknesses 175 and 165 nm were epitaxially prepared on $\rm{SrTiO}_3$(100) and $\rm{NdGaO}_3$(110) substrates by %off-axis
pulsed-laser deposition using stoichiometric targets. The c direction of the LMO film is orthogonal to the surface normal. In the A-type antiferromagnet LMO, the Mn spins order ferromagnetically in the ab plane while they are stacked antiferromagnetically along the c direction. With the electric-field vector parallel to the ab plane we thus excited and probed ferromagnetic spin correlations in this experiment. The Neel and Curie temperatures of the LMO and LCMO films are $\mathrm{T}_\mathrm{N}\sim$ 150 K and $\mathrm{T}_\mathrm{C} = 256$ K, respectively. The metal-insulator transition temperature of the latter is $\mathrm{T}_{\mathrm{MI}} = 264$ K \cite{Walter}.\newline%
\section{Results}
\begin{figure}
\includegraphics[width=0.9\columnwidth]{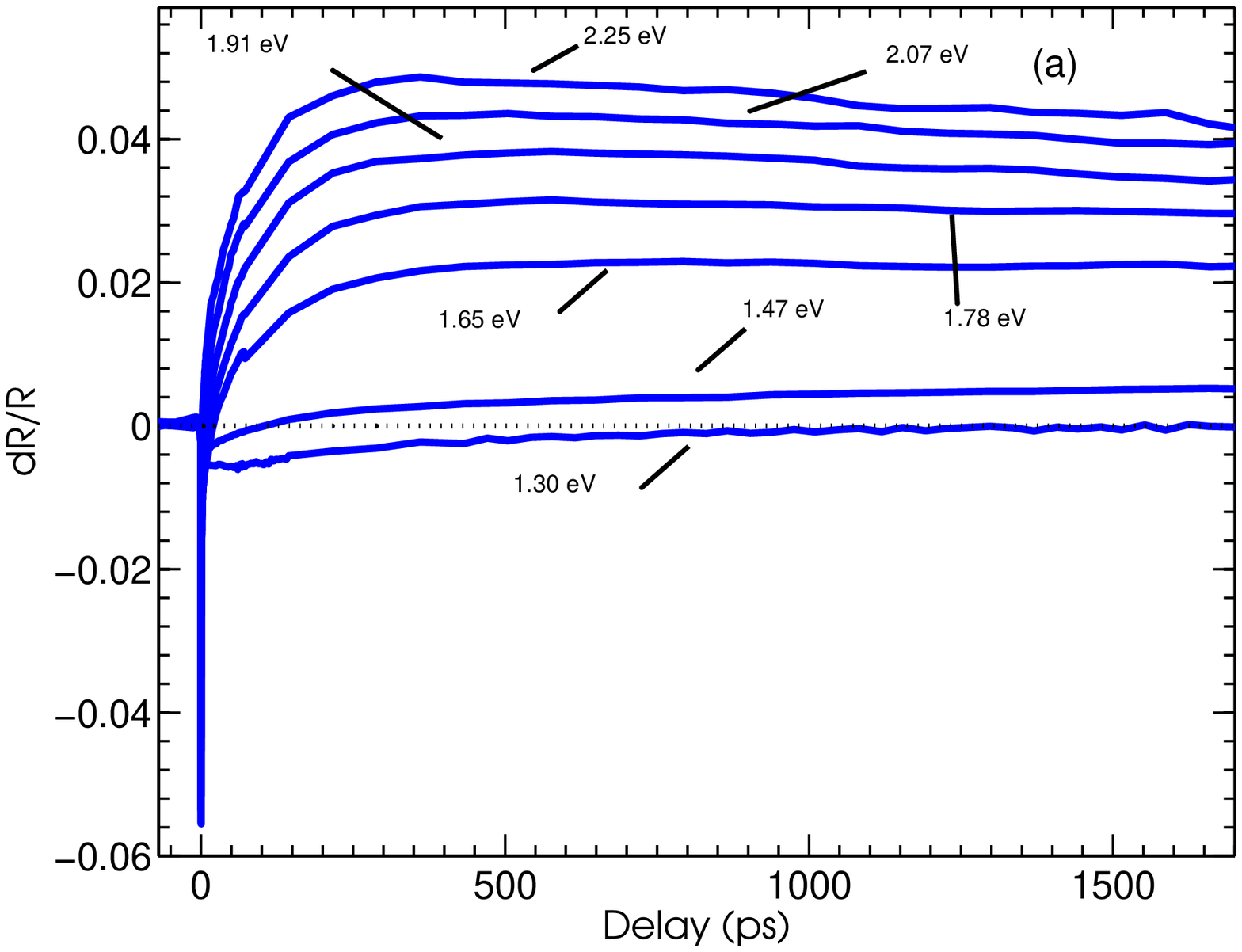}
\includegraphics[width=0.9\columnwidth]{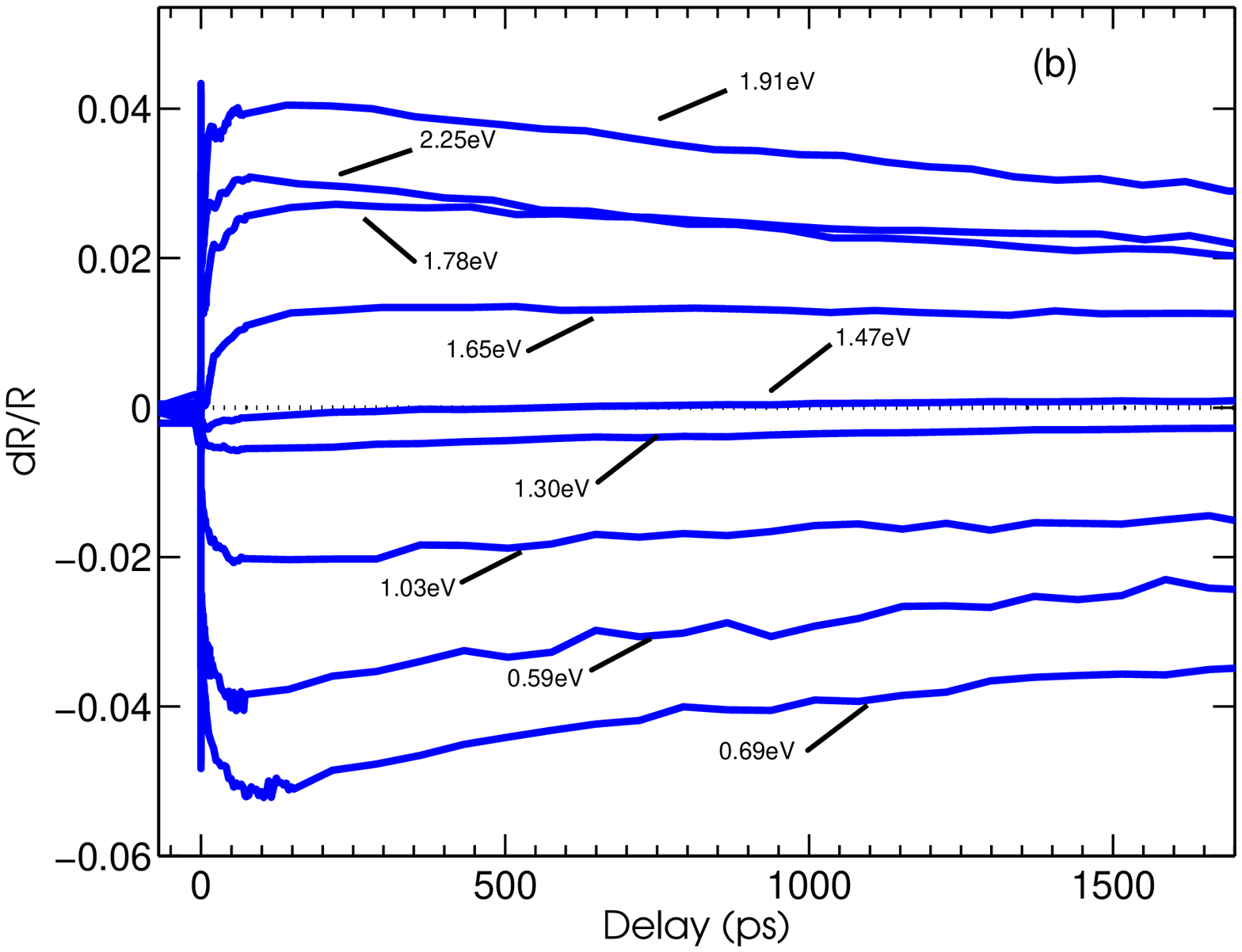}
\caption{(Color online) Transient reflectivity in (a) LMO and (b) LCMO for different probe energies when pumping with 1.6 eV. Notice in particular the change in the final relaxation time when changing probe energy and the zero crossing ($\sim100$ ps, LMO and $\sim 550$ ps LCMO) when probing with 1.47eV, features that are so far unexplained by the literature.}
\label{data}
\end{figure}

The reflectivity transients of LMO between 1.31 and 2.25 eV at 80K are shown in Fig. \ref{data}(a). After the initial negative peak at t = 0 the transient increases, changes sign between $\sim$5 ps (Epr = 2.25 eV) and $\sim$100 ps (1.48 eV), reaches a maximum between $\sim$400 ps (2.25 eV) and beyond 2ns (1.48 eV), and eventually decays exponentially on a ns time scale.
We model the experimental results using a modified three temperature model. The standard three temperature model \cite{Beaurepaire} describes the energy flow into and from the electron, lattice and spin subsystem respectively, followed by a usually very slow heat diffusion component:
 \begin{eqnarray}
 dR/R= A_e e^{-t/\tau_{el}}+ A_l\left(1-e^{-t/\tau_{el}}\right)e^{-t/\tau_{ls}}\nonumber \\ + A_s\left(1-e^{-t/\tau_{ls}}\right)e^{-t/\tau_{d}}
 \label{eq:standard}
 \end{eqnarray}
 where $A_{e,l,s}$ describes the magnitude on $dR/R$ arising from an increased temperature in the electron, lattice and spin system respectively. Similarly, $\tau_{d, el, ls}$ denotes the time constants for heat diffusion (ns-scale), energy transfer between the electron and lattice systems (ps-scale) and lattice and spin systems respectively (100 ps scale). A different interpretation for the ns-scale process has been suggested\cite{Ren} where the slowest relaxation back to the ground state was attributed to metastable quasiparticle dynamics involving spin-flip processes. As the slowest component we observe is present also in the paramagnetic room temperature phase (not shown here) we keep the orthodox heat-diffusion interpretation.
 The described model should be viable as long as only thermal processes occur. It has been previously reported in similar materials that this is the case after $\sim 1 $ps, after which the initial non-equilibrium electron gas has had time to thermalize back to a fermi-distribution. It should be noted that the assumption of thermal processes implies that the time constants for the energy transfers between the subsystem as well as the subsequent heat diffusion will be the same regardless of probing energy.
 \begin{figure}
\includegraphics[width=0.9\columnwidth]{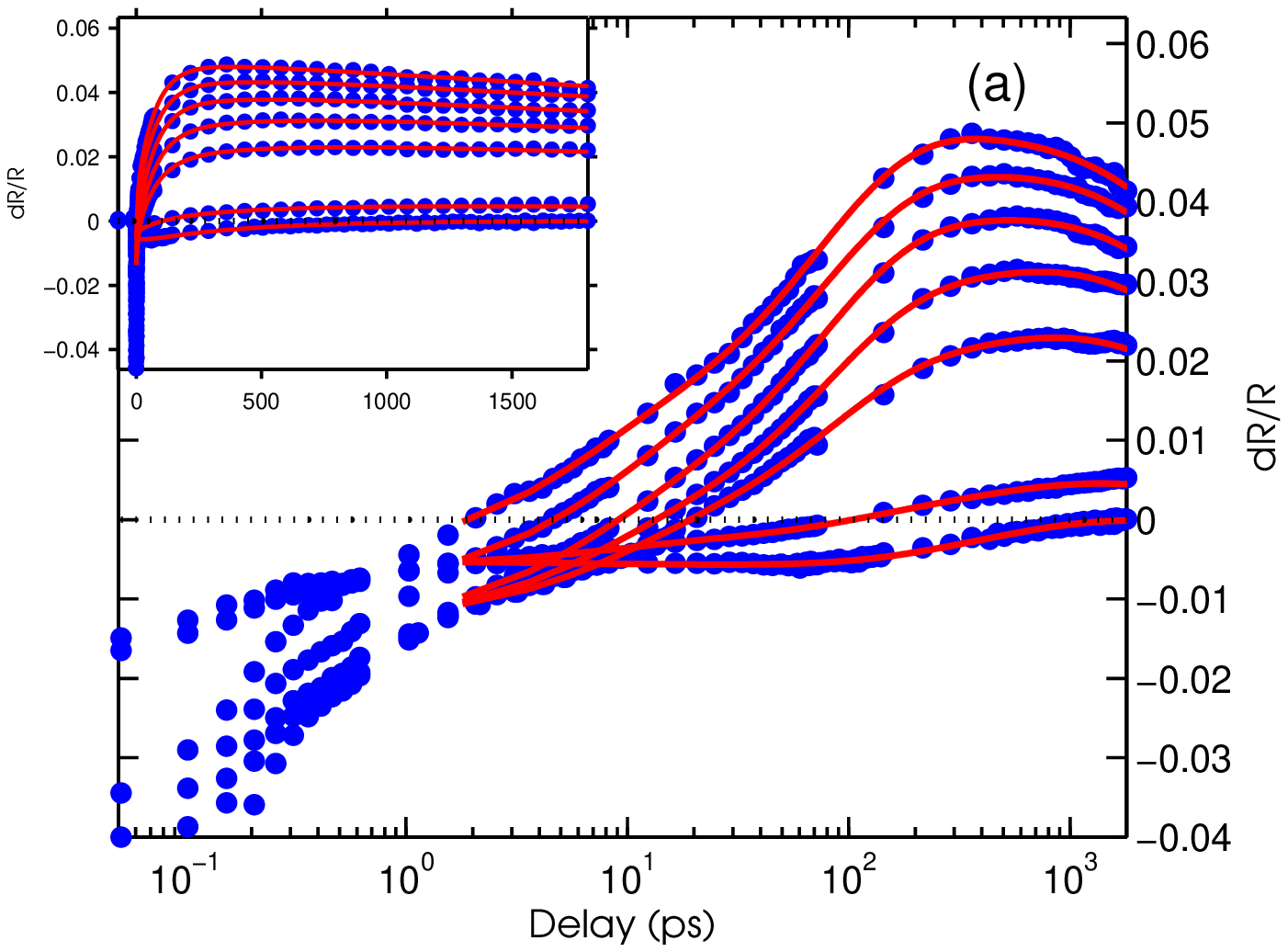}
\includegraphics[width=0.9\columnwidth]{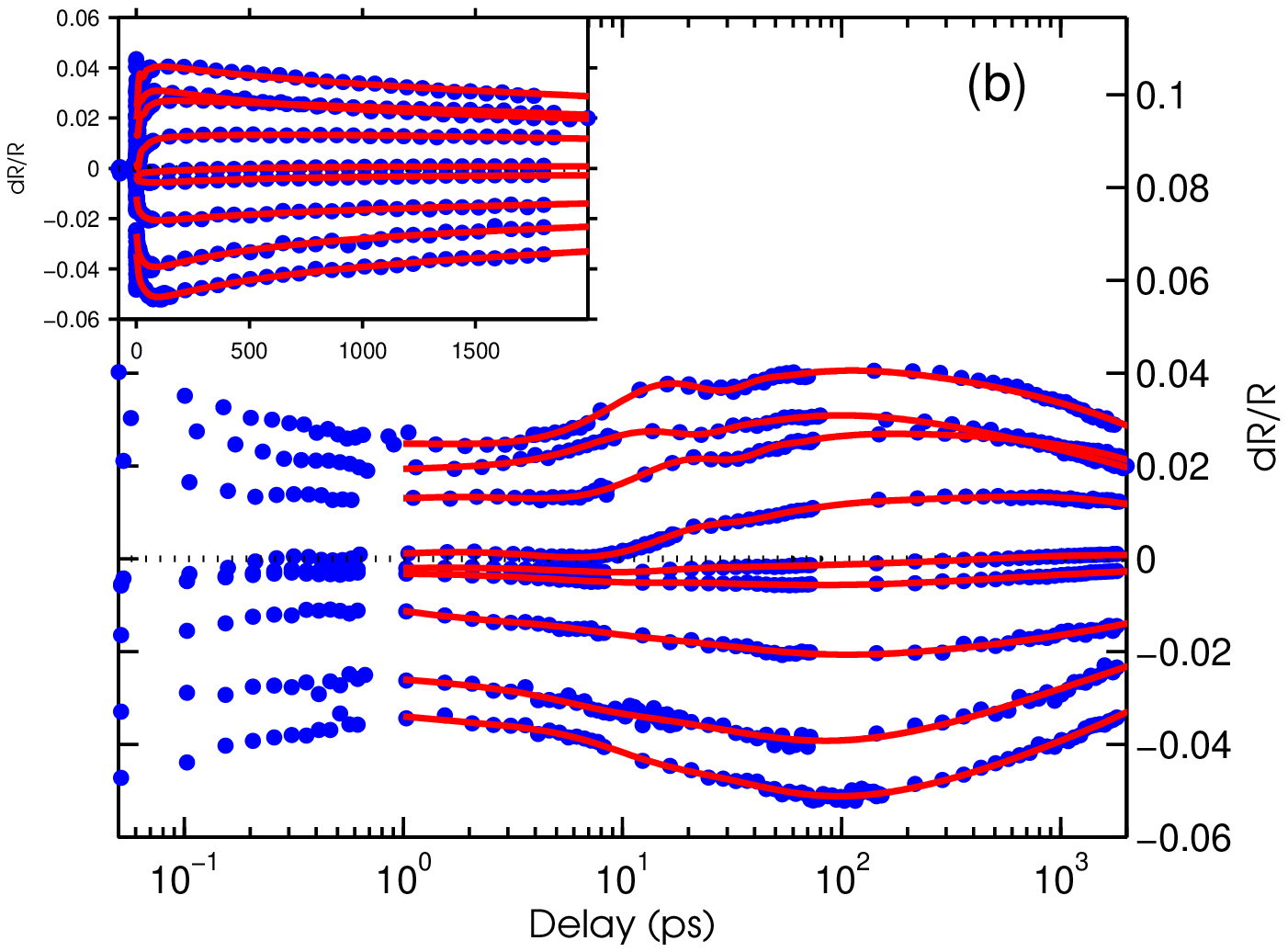}
\caption{(Color online) The computed fittings to the experimental (a) LMO and (b) LCMO data plotted on a semi-log scale. The dots represent the experimental data while the lines is the corresponding fittings. The inset show the same fittings using a linear scale. The transients shown are the same as in Fig. \ref{data}.}
\label{fit}
\end{figure}
 A systematic fitting procedure, in which the time constants were held constant between all probe energies, were performed on all our data. It is possible to fit individual transients using models with fewer components, but this simplification comes with the expense that the thermal interpretation with time constants independent of probing energy has to be abandoned. With this systematic fitting procedure it is possible to fit all LMO data using an e-l time constant $\tau_{el}= 10$ ps and a l-s time constant $\tau_{ls}=75$ ps. Our value of $\tau_{el}$ is in good agreement with previous differential transmission studies, though no sign of the spin dynamics that we link to the $A_s$ component was reported\cite{Tamaru}. In addition, a consistent fit to our data requires two slow components $\tau_{d1}=720$ ps and $\tau_{d2}=8960$ ps:
   \begin{eqnarray}
 dR/R=A_e e^{-t/\tau_{el}}+ A_l\left(1-e^{-t/\tau_{el}}\right)e^{-t/\tau_{ls}}\nonumber \\ + A_{s}\left(1-e^{-t/\tau_{ls}}\right)\left[e^{-t/\tau_{d1}}+C_r e^{-t/\tau_{d2}}\right]
 \label{eq:model}
 \end{eqnarray}
 where $C_r$ is the relative weight between the two slow components. The necessity of both these components can be directly deduced from Fig. \ref{data}(a) by looking at the $1.3$ and $1.47$ eV data. First of all, the differential reflectivity seen by the $1.3$ eV probe shows a, in comparison, very fast decay to zero signal, clearly not compatible with what is seen at higher probe energies. Also, the $1.47$ eV transient has its maximum after a delay larger than 1500 ps, indicating at least two components at interplay after several hundreds of picoseconds.%zero-crossing after $~100$ps, indicating at least two components in interplay at the $100$ps scale.

\begin{figure}
\includegraphics[width=0.9\columnwidth]{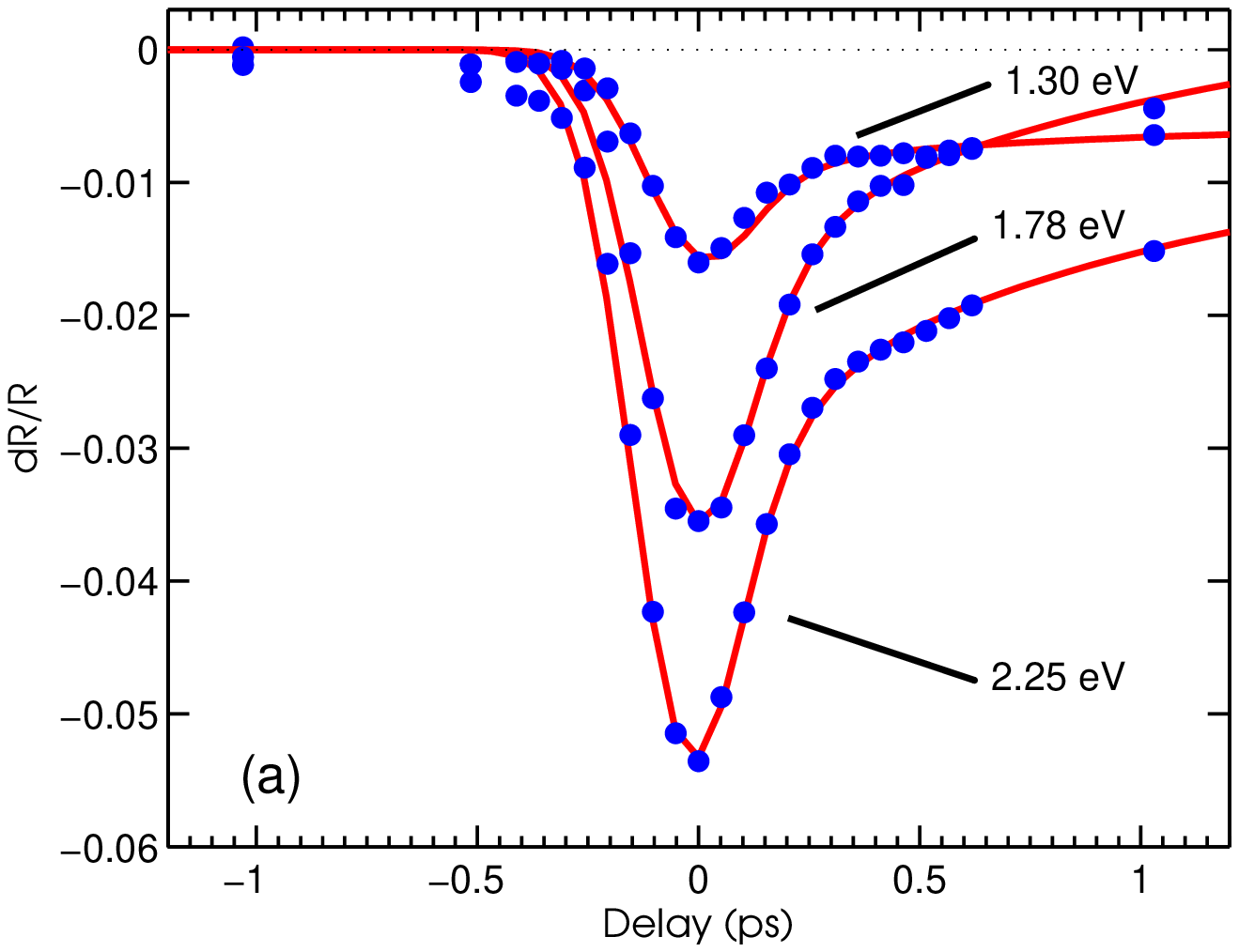}
\includegraphics[width=0.9\columnwidth]{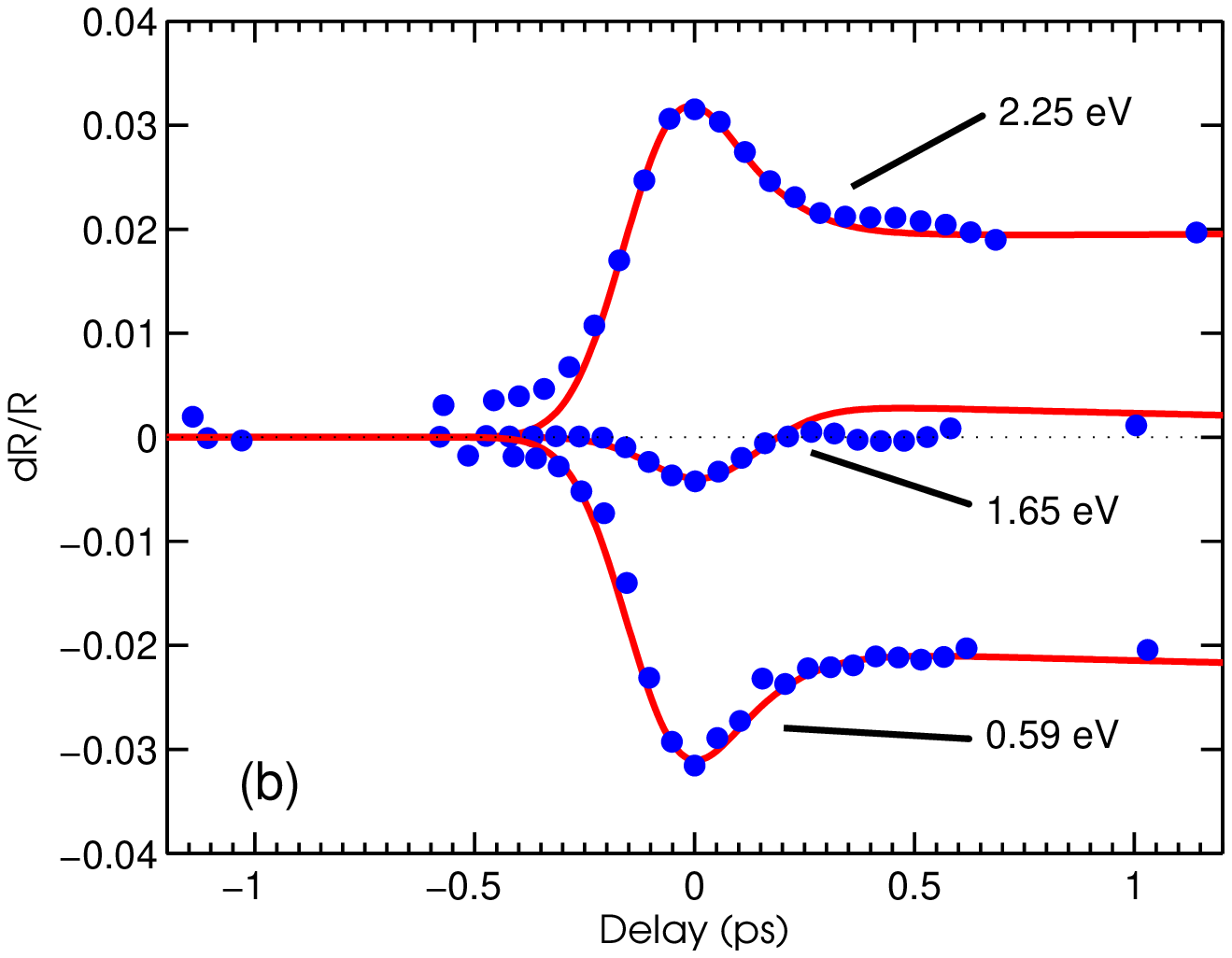}
\caption{(Color online) Sub-picosecond transients in LMO (a) and LCMO (b).  In both samples we see a roughly gaussian peak followed by very fast ($< 1$ ps) decay processes, while only LCMO shows signs of coherent optical phonons. The blue dots represents the experimental data while the red lines are the corresponding fittings.}
\label{fig:Peak}
\end{figure}

The corresponding reflectivity transients in Ca-doped LCMO at 80 K are displayed in Fig. \ref{data}(b). Apart from $\mathrm{E}_{pr} = 1.65$ and $1.47$ eV, where the sign changes around 0.6 (see Fig. \ref{fig:Peak}(b)) and 550 ps, respectively , the initial peak and long-lived transient show the same sign depending on $\mathrm{E}_{pr}$.  Note that above $t\sim 1$ ps, the transient increases or decreases much faster than in LMO, reaching an extremum between about 70 ps (2.25 and 1.31 eV) and roughly 400 ps (1.65eV) before exponentially decaying within several ns. Additionally, the transient is modulated by a damped oscillation below $\sim 40$ ps.
The same modified three-temperature model introduced above, Eq. (\ref{eq:model}), is sufficient also to describe the low-temperature transient reflectivity of LCMO, with the addition of an oscillating part attributed to the release of coherent acoustic phonons \cite{Thomsen}.
The resulting fit is shown in Fig.\ \ref{fit}(b) and gives for LCMO the time constants $\tau_{el}=2.6$ ps, $\tau_{ls}=31.5$ ps, $\tau_{d1}=475$ ps, $\tau_{d2}=7000$ ps and an oscillation period $E_{pr}\tau_p  \sim 60 $eV ps. This indicates a faster response than in LMO, which may be attributable to the stronger correlations in LCMO. Our observed time constants are consistent with the values obtained from time resolved conductivity measurement \cite{Averitt} and time resolved MOKE measurements \cite{McGill, OgasawaraKerr, OgasawaraGeneral},  although strain effects due to the use of different substrates might influence the exact numbers.

\begin{figure}
\includegraphics[width=0.45\columnwidth, angle=270]{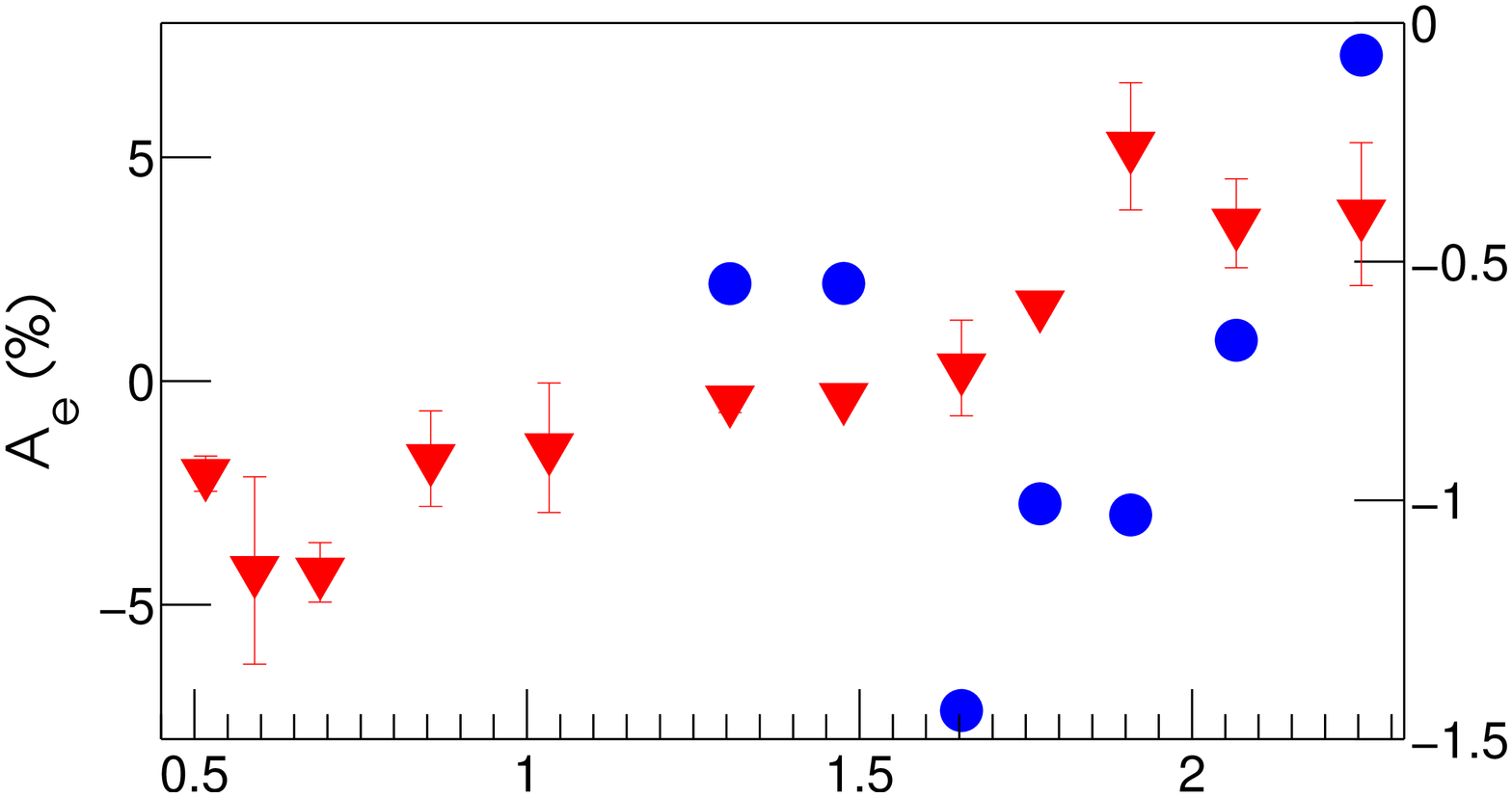}
\includegraphics[width=0.45\columnwidth,angle=270]{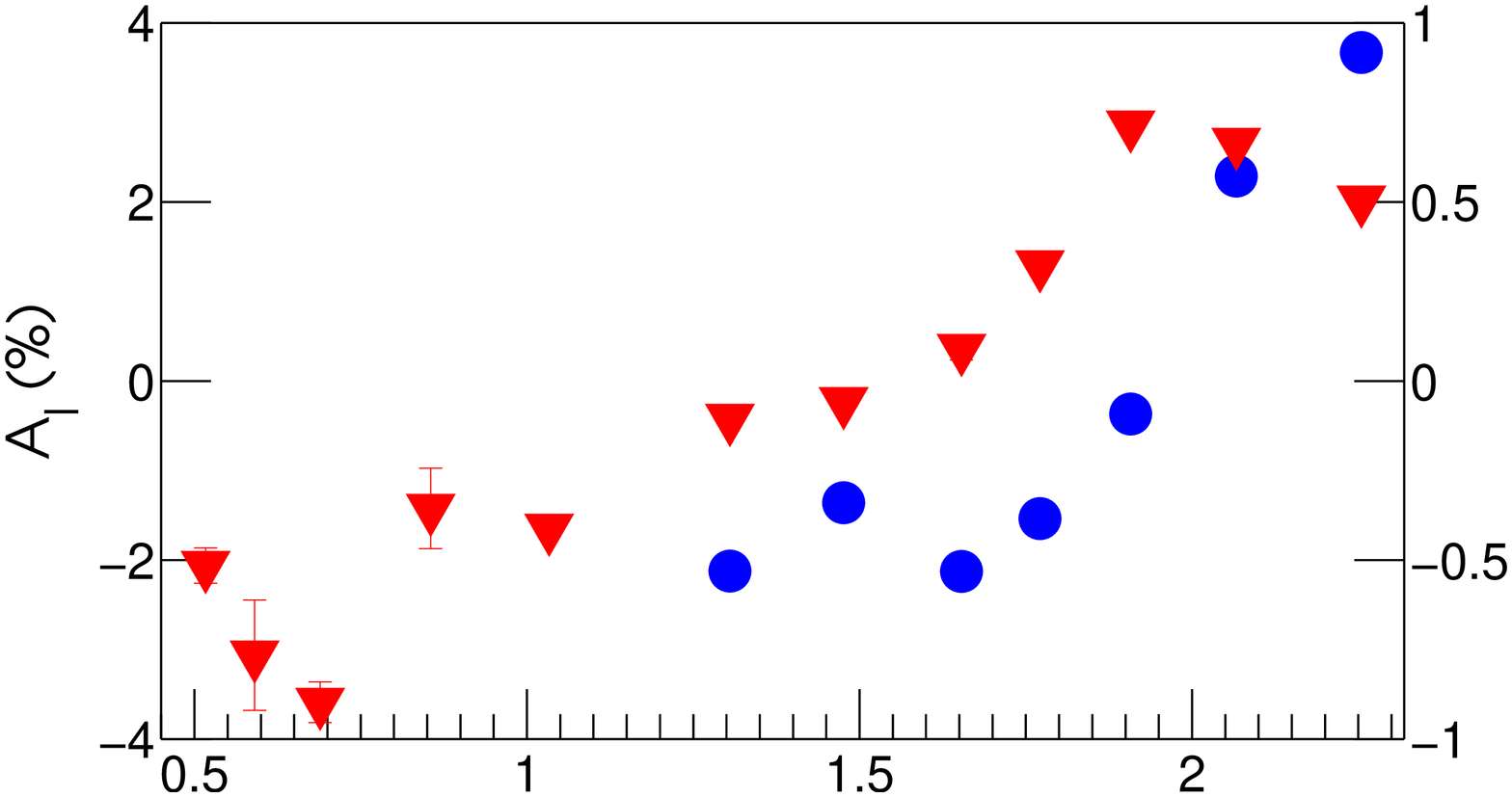}
\includegraphics[width=0.45\columnwidth,angle=270]{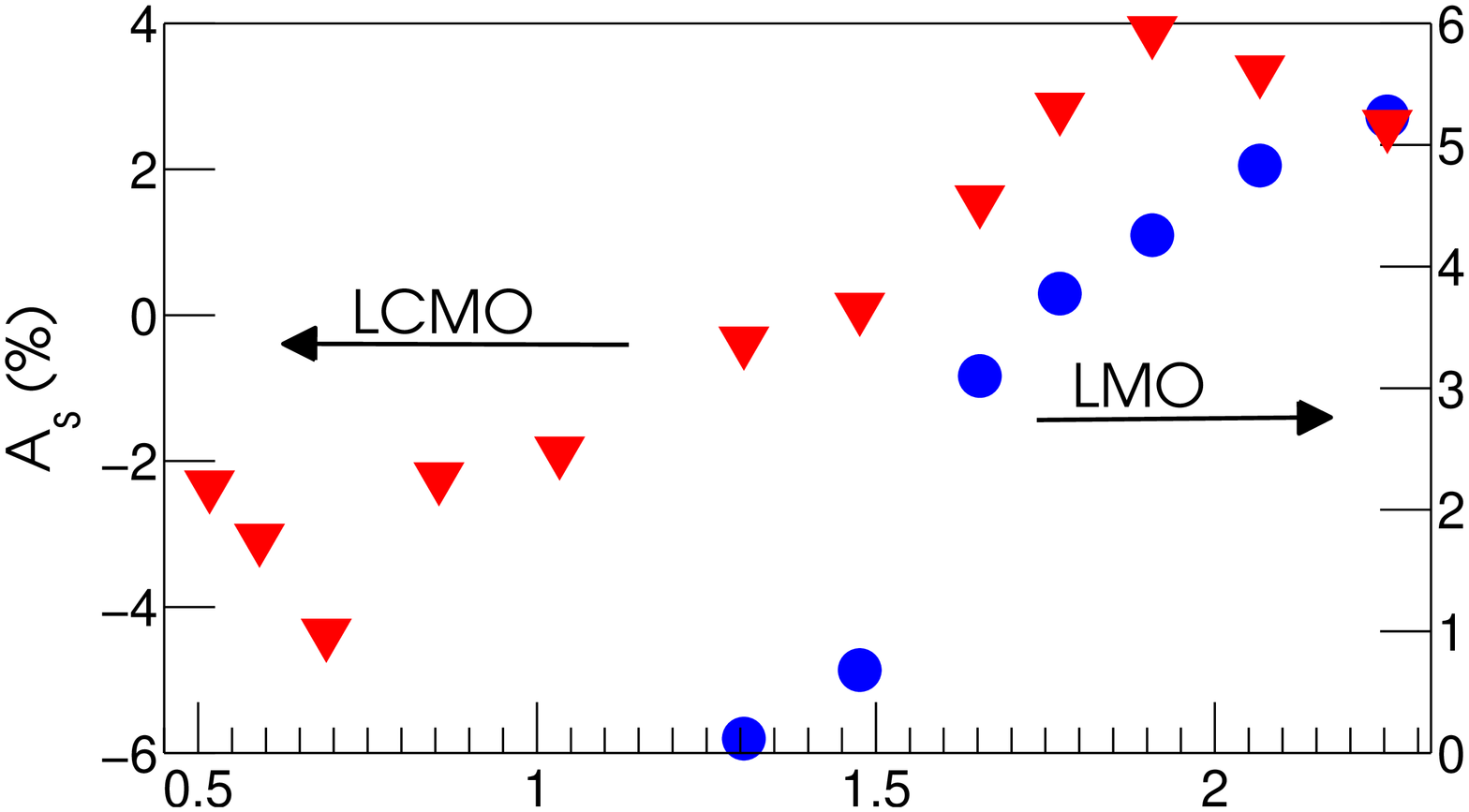}
\includegraphics[width=0.49\columnwidth,angle=270]{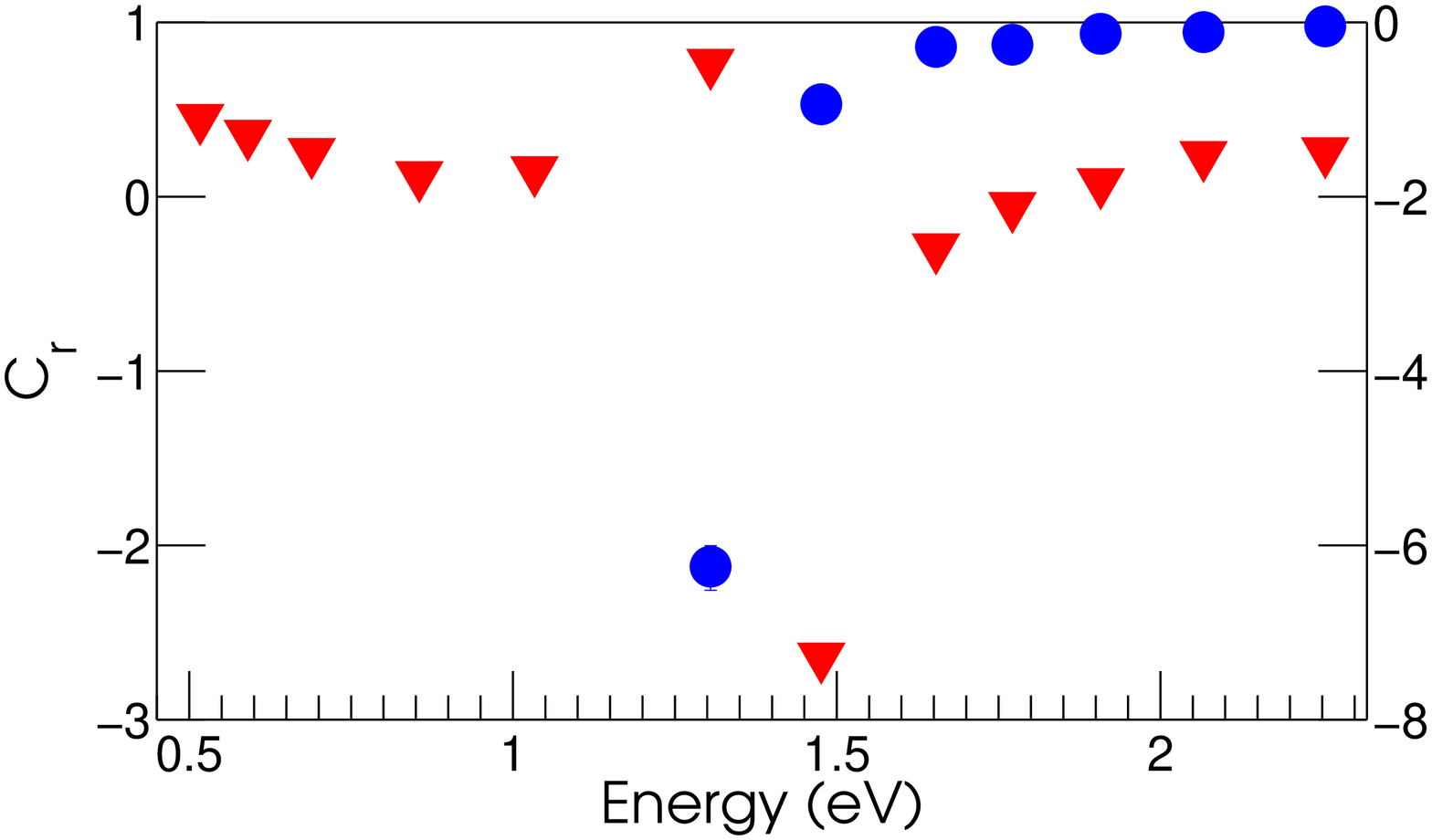}

\caption{(Color online) The computed fitting parameters of LMO (blue circles, right axis) and  LCMO (red triangles, left axis) where the errorbars indicates 95\% confidence intervals. }
\label{parameters}
\end{figure}

In Fig. \ref{parameters} we show the values of the magnitudes $A_{e,l,s,}$ and $C_r$ obtained for LMO and LCMO from the fitting procedure described above. $A_e$, $A_l$ and $A_s$ all show very similar trends with probe energy, while the dependence of $C_r$ on probe energy is very different. The accuracy of the parameters is best for $A_s$ and $C_r$ due to the large signals and decoupling in time from the other processes. In LCMO, accurate determination of $A_e$ and $A_l$ is hindered by the simultaneous presence of oscillations from the coherent phonons, something that is reflected in the large errorbars for these parameters. In both LMO and LCMO, $C_r$ shows a divergent behaviour. The divergence occurs  when approaching the energies $1.2$eV (LMO) and $1.4$eV (LCMO) and is discussed below in Section \ref{sec:Discussion} for the insulating LMO, while so far the divergence in LCMO is unexplained.

 Having successfully modeled the transients above the picosecond scale with the three temperature model, we turn now to the non-thermal sub-picosecond  response. Typical ultrafast transients for LMO and LCMO is shown in Fig. \ref{fig:Peak} (a) and (b) respectively.  The main shape of the peak is gaussian with a width corresponding to the cross-correlation between the pump and probe pulses. Convoluted with the gaussian shape is a fast decay due to the thermalization of the excited electrons, while LCMO also shows high frequency oscillations with a period $\sim 600$fs, possibly from strongly damped coherent optical phonons\cite{Lim} . To account for these features, neglecting the oscillations, we model the number of electrons $N$ in an excited state as
\begin{equation}
\frac{\mathrm{d}N}{\mathrm{d}t}=P(t)-N/\tau_t
\label{eq:peak}
\end{equation}
where $\tau_t$ is the decay rate (thermalization time) and $P(t) \sim \exp\left(-t^2/\tau^2_p\right)$ is the source term with a width $\tau_p$.  Solving Eq. (\ref{eq:peak}) and assuming further that $dR/R\sim N$, the non-thermal contribution to the transient response gets the form
\begin{equation}
dR/R=\frac{1}{2}\left[1-\mathrm{erf}\left(\tau_p/2\tau_t-t/\tau_p\right)\right] A_t e^{-t/\tau_t}
\label{eq:peakfit}
\end{equation}
with erf the error function and $A_t$ a parameter to be determined, together with $\tau_t$, by fitting to the data.  Requiring the fittings in Eq. (\ref{eq:model})  and (\ref{eq:peakfit})  to be consistent, a single equation describing the complete transient response is obtained
\begin{eqnarray}
dR/R=\frac{1}{2}\left[1-\mathrm{erf}\left(\tau_p/2\tau_t-t/\tau_p\right)\right] \times \nonumber \\ \left[ A_t e^{-t/\tau_t}+A_{e}\left(1-e^{-t/\tau_{t}}\right)e^{-t/\tau_{el}}+\dots\right]
 \label{eq:complete}
 \end{eqnarray}
where $\dots$ denotes the remaining part of Eq. (\ref{eq:model}).

When fitting the peak and subsequent carrier thermalization the duration $\tau_p$ of the pump pulse is assumed to be the nominal 170 fs. Using the parameters obtained from Eq. (\ref{eq:model}), there are two parameters, $\tau_t$ and $A_t$, to be determined by the same fitting procedure as above. Using this scheme, the thermalization time $\tau_t$ is found to be $67$fs for LMO and $78$fs for LCMO. The amplitude $A_t$, describing the impact of the excitations of the free carriers, exhibits an energy dependence
similar to that of $A_e$, $A_l$ and $A_s$.

The response modeled in Eq. (\ref{eq:complete}) is completely adequate for LCMO but for LMO an additional fast component with decay time $850$ fs is needed for a satisfactory fitting. The existence of a two component decay of photoexcited carriers in Mott insulators have been reported previously\cite{Perfetti} and has been attributed to an immediate response following the photo induced insulator-metal transition followed by a transition to a slower relaxation after the recovery of the insulating bandgap. %In this view it is not surprising that the thermalization of the free carriers towards a Fermi distributed electron gas in LMO and LCMO occurs at such similar time scales, as both samples behaves at metals.

\section{Discussion}
\label{sec:Discussion}
In pump-probe spectroscopy, electric-dipole allowed transitions are excited and probed. As their strength depends on the densities of initial and final states (DOS) and the transition probabilities between them, reflectivity transients arise from either DOS changes or transition-probability changes.
From numerous studies of the equilibrium optical properties \cite{Grenier, Kovaleva, Rauer} and band-structure calculations \cite{Feiner} of the manganites two kinds of electric-dipole allowed transitions may be expected in the spectral range $0.8-2.3$ eV: transitions between the 3d orbitals of neighboring Mn ions (d-d) and transitions between O2p and Mn3d states (p-d). While the former are highly sensitive to the relative orientation of Mn spins and thus on temperature, the latter are expected to be temperature independent\cite{Feiner} .

While the initial photoexcitation by the pump pulse might involve both d-d and p-d transitions, the strong temperature dependence of the intermediate and slow transient components, especially across $\mathrm{T}_\mathrm{N}$ and $\mathrm{T}_\mathrm{C}$, clearly indicates that these components arise from probing d-d transitions \cite{Lobad, OgasawaraKerr, Kovaleva, Rauer}.

Let us summarize the current understanding of photoinduced dynamics in LMO or LCMO. The 1.6 eV photoexcitation creates pairs of electrons and holes (quasiparticles).
Within $< 1$ps, the electrons excited into the upper d band thermalise to a Fermi distribution at an elevated temperature $T_e$, out of equilibrium with the lattice and the spin system, occupying formerly empty states at the lower band edge \cite{OgasawaraGeneral}. We found that this process is a two component process in the insulating LMO ($\tau=67$ and $850$ fs respectively), while a single component is sufficient for LCMO ($\tau=78$)fs.
Subsequently, the quasiparticles recombine via broad-band phonon emission (e-l relaxation).
The lattice temperature rise at the expense of the electron temperature.
In the next step, any excess lattice energy is transferred to the spin system (l-s relaxation) via spin-orbit interaction, the lattice and spin temperatures equilibrate \cite{OgasawaraGeneral, Beaurepaire}. This effect was observed in CMR manganites before, the s-l relaxation time was found to be $\sim 20$ to $200$ ps, depending on temperature \cite{Lobad}. %\newline%%%%%%%%%%%%%%%%%%%%%%5

At this point, the temperatures of all three baths are in equilibrium. The temperature in the illuminated volume decreases via heat diffusion. We showed above that this heat diffusion consist of (at least) two components. To interpret these two processes we turn to LMO. The relevant band in LMO that is probed and pumped in the visible range is the high spin $\rm{e}_{\rm{g}}- \rm{e}_{\rm{g}}$ charge transfer between adjacent $\rm{Mn}^{3+}$ ions via the bridging $\rm{O}^{2-}$ ion. This charge transfer is mediated both by super-exchange and double-exchange, hence being very sensitive to the $\rm{Mn}-\rm{O}-\rm{Mn}$ bond angle, indicating the importance of the static Jahn-Teller distortion. Thus the spin-temperature dynamics become visible because the probed high spin d-d transition acts as a thermometer: spin disorder results in a weakening of the transition probability, while the lattice dynamics become visible because an increased lattice temperature causes a relaxation of the Jahn-Teller distortion which thereby changes the $\rm{Mn}-\rm{O}-\rm{Mn}$ bond angle. The optical absorption in this band decreases abruptly around $1.5$ eV \cite{Tobe}, i.e with $E_{pr}=1.3$ eV LMO is more transparent, indicating that the film-substrate interface is more important.

 Hence we attribute the fast heat diffusion process to the film-to-substrate diffusion, a conclusion shared with reference \cite{Tamaru}. Analogously, with higher $E_{pr}$ LMO is more opaque, giving us reason to attribute the very slow relaxation process to the traditional in-film heat diffusion in accordance with previous assumptions \cite{Liu, OgasawaraGeneral}.

\begin{figure}
\includegraphics[width=0.95\columnwidth]{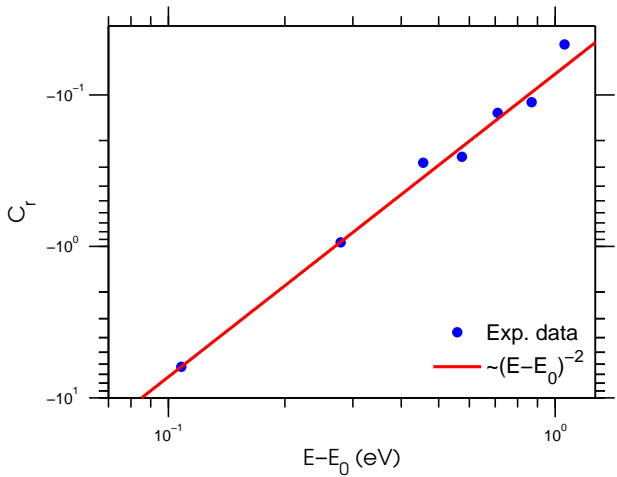}
\caption{(Color online) The measured coefficient $C_r$ compared to the theoretical curve $(E-E_0)^{-2}$ on a log-log scale. }
\label{Cr}
\end{figure}

 To further motivate this interpretation, consider the coefficient $C_r$. The magnitude of this coefficient should relate inversely to the number of high spin d-d excitations probed at a given energy. As LMO is an indirect semiconductor\cite{Olle},  the probability of such an excitation, assuming parabolic density of states close to the optical gap, should be proportional to $(E-E_0)^2$ where $E$ is the probe energy and $E_0$ is the energy of the band gap plus the mediating phonon\cite{Marder}. Therefore we expect $C_r \sim (E-E_0)^{-2} $. In Fig. \ref{Cr} we have plotted $C_r$ as a function of energy together with  $(E-E_0)^{-2}$. $E_0$ was obtained by fitting the data to the model, resulting in $E_0=1.195$eV. This value should be compared to the previously published  value $\mathrm{E}_\mathrm{gap}\sim 1.2$eV\cite{Tobe}.  The good agreement between the value obtained here with the previously published value, together with the good quality of the fit, shows that the film-to-substrate interpretation is reasonable.

A similar analysis can be made for LCMO, but because LCMO is metallic at 80K, the lack of a band-gap never causes us to expect a diminishment in relative intensity of the in-film heat diffusion when lowering $E_{pr}$. Hence, the film-to-substrate component will only be explicit near a sign change, as is the case for $E_{pr}=1.47$ eV, see Fig. \ref{data}(b). Keeping the above discussion in mind, we can understand the absence of LMO spin dynamics in previous differential transmission studies\cite{Tamaru} as due to the thickness of the thin films used there ($70$ nm), thus increasing the effect of the film-to-substrate diffusion for all probe wavelengths.

%Due to the time scale of this diffusion process, it shadows the spin dynamics.
%\newline
In summary, we have performed resonance pump-probe studies on the femtosecond scale on both the undoped $\rm{La}\rm{Mn}\rm{O}_3$ and the doped $\rm{La}_{0.7}\rm{Ca}_{0.3}\rm{Mn}\rm{O}_3$. In spite of having very different properties, with LMO being an insulator and LCMO being a metal at 80K, the observed differential reflectivity transients are very similar in character. We have developed one general mathematical model, Eq. (\ref{eq:complete}), which captures all observed phenomena in the transients, including free carrier, electron, lattice and spin dynamics. This model differs from the previously used in that it includes two different heat diffusion channels, in-film diffusion and film-to-substrate diffusion. We conclude that the penetration depth is generally large enough for the film-substrate interface to have an effect, and that this effect can be used to deduce the bandgap in semiconducting samples.

%\bibliography{refs}

\end{document}